# Fieldoscopy at the Quantum Limit

Dmitry A. Zimin[1,2], Arjun Ashoka[1], Florentin Reiter[3,4] and Akshay Rao[1]

[1]*Cavendish Laboratory, Department of Physics, Cambridge University, CB3 0HF Cambridge, United Kingdom*
[2]*Laboratory of Physical Chemistry, ETH Zürich, Vladimir-Prelog-Weg 2, 8049, Zürich, Switzerland*
[3]*Institute for Quantum Electronics, ETH Zürich, 8093 Zürich, Switzerland*
[4]*Quantum Center, ETH Zürich, 8093, Zürich, Switzerland*

**Abstract**

We demonstrate a novel concept for measuring time-varying electric field transients of petahertz-scale photons down to a single-photon regime. We observe a clear transition from classical to quantum nature of light that agrees with our Monte Carlo model. We reach unprecedented yoctojoule-level sensitivity ($10^{-24}$ J) and a dynamic range exceeding 90 decibels. We utilize this capability to measure time-dependent intrapulse light coherence - a regime inaccessible to conventional, time-averaged spectroscopy. This opens new avenues for quantum information, cryptography, and quantum light-matter interactions on sub-cycle time scales with attosecond precision.

# Introduction

Light coherence describes the relationship between the fluctuations of photons in a light source and their degree of distribution in time and space [1]. This property is fundamental for distinguishing laser light [2] from incoherent sources and has enabled widespread application across disciplines, from fundamental physics [3, 4], precision measurements [5, 6], industrial manufacturing [7], data storage [8] or medical diagnostics [9, 10, 11]. Coherence enables ultrafast dynamics of matter to be probed using femtosecond lasers, offering insights into chemical reactions [12, 13] and material dynamics [14]. By measuring the electric field of light, coherent processes can be studied on sub-cycle time scales with attosecond precision and unprecedented sensitivity [11, 15]. This concept, known as fieldoscopy [16] has recently been utilized to measure the formation of the optical response [15], ultrafast magnetism [17], and the light-matter energy transfer [18] in photoexcited solids.

Coherence is directly linked to the quantum properties of light, i.e. photon statistics, making it a crucial quantity in quantum optics and quantum information science. For example, weak coherent states of a laser with a low mean photon number are widely utilized in quantum cryptography to create single-photon pulses. Light coherence and photon statistics however can be affected by nonlinear effects [19]. In a laser, self-phase modulation, gain saturation and recovery [20], amplified spontaneous emission [21], intracavity soliton dynamics [22] or mode-lock instabilities [23] are some of the examples. In pulsed lasers (e.g. mode-locked lasers) the electric field strength varies across the pulse which inevitably leads to the temporal dependence of coherence, i.e. intrapulse coherence. While time-averaged pulse coherence can be measured [24], the coherence variation within a pulse (intrapulse coherence) requires a direct access to the electric field in the regime dominated by the quantum nature of light.

Fieldoscopy enables direct access to the electric field of light and has seen rapid developments in recent years [25, 26]. High-frequency photons are particularly advantageous for generating single-photon pulses, as the substantial amount of energy can be concentrated in a single quantum. Petahertz fieldoscopy typically requires intense, few-cycle carrier-envelope phase (CEP) stable pulses, complex experimental infrastructure, and highly non-linear light matter interaction [27, 28, 29]. This complexity limits the sensitivity required to access the single-photon regime, where the quantum nature of light is dominant.

Here we perform the first field-resolved measurements in the single-photon regime. We measure 1.2 eV (193 zJ, 1030 nm) photons with the average pulse energy down to 876 yoctojoules – a regime where 99.55 % of pulses are in a vacuum state and out of the remaining 0.45 % pulses, 99.78 % are in a single photon state. Our approach does not rely on complex laser infrastructure, highly nonlinear light–matter interactions, or carrier-envelope phase (CEP)-stable pulses. We observe a distinct transition from the classical to the quantum regime, which is well captured by our Monte Carlo model (more in Methods). Furthermore, we employ this concept to probe the time-dependent intrapulse coherence of a mode-locked laser, revealing how coherence varies across the pulse. We find that the more intense regions of the pulse exhibit reduced coherence compared to the weaker ones. We attribute this to the presence of an intensity-dependent decoherence process which admixes a thermal component to the ideal coherent state.

Our work extends petahertz fieldoscopy into the quantum regime opening new possibilities for sub-cycle quantum information, cryptography, and precision metrology. The ability to directly access the electric field of weak states of light represents a paradigm shift in quantum optics, providing a new tool for investigating coherence dynamics at sub-cycle timescales and paving the way towards quantum lightwave electronics.

**Results**

The concept, schematically shown in Figure 1a, is based on the recently demonstrated generalized heterodyne optical sampling technique (GHOST) [30]. This technique is similar to the well-established electro-optic sampling [31, 32] which is extensively utilized [33, 34] to sample the electric field of light in the terahertz domain. In these schemes, two pulses, the sampling pulse and the test pulse (whose electric field is of interest) with a temporal delay $\tau$,

$$E_S(t) = \bar{E}_S e^{i(\omega_S t + \varphi_S)}, \tag{1}$$

$$E_T(t,\tau) = \bar{E}_t e^{i(\omega_T t + \varphi_T - \omega_T \tau)}, \tag{2}$$

interact in a nonlinear crystal to produce a nonlinearly generated field (NP) that encodes the test-field's instantaneous value:

$$E_{NP}(t,\tau) \approx \bar{E}_S^{n-1} \bar{E}_T e^{i[(((n-1)\omega_S + \omega_T)t + (n-1)\varphi_S + \varphi_T - \omega_T \tau)]}. \tag{3}$$

Here, $\bar{E}_S$ and $\bar{E}_T$ are field strengths, $\omega_S$ and $\omega_T$ are frequencies while $\varphi_S$ and $\varphi_T$ are the phases of the sampling and test fields. The NP can be any nonlinear mixing signal [30], such as sum-frequency (SFG), difference-frequency (DFG) generation, four-wave mixing (FWM), etc. By interfering the nonlinearly generated field with a "stationary" local oscillator (LO):

$$E_{LO}(t) \approx \bar{E}_S^m e^{i(m\omega_S t + m\varphi_S)}, \tag{4}$$

one can retrieve the electric field of the test pulse at each delay. The LO can be simply the sampling field itself or a nonlinearly generated light, e.g. second, third, fourth or higher harmonics [30]. The $n$ and $m$ factors correspond to the light-matter interaction orders used for LO and NP, respectively. The intensity of the interference I($\tau$, $\omega_d$) at the detection frequency $\omega_d$ is measured with a photodiode and is proportional to the electric field of the test pulse at time $\tau$. Experimentally, $\omega_d$ is set by the bandpass filter (BP),

$$I_{LO+NP}(\tau, \omega_d) \approx \bar{E}_S^{m+n-1} \bar{E}_T \cos(-\omega_d \tau - m\varphi_S + (n-1)\varphi_S + \varphi_T). \tag{5}$$

As we show further, unlike electro-optic sampling where $m = 1$, $n = 2$, and previous implementations of GHOST ($m = 3$, $n = 2$ [30], or $m = 2$, $n = 3$ [35]) balanced $m = n$ combination provides a previously unexplored opportunity. In the unbalanced case, $I_{LO+NP}(\tau, \omega_d) \approx \varphi_T - (m - n + 1)\varphi_S$ (Eq. 5). If the detection is not performed in a single-shot, unevenly amplified instabilities of $\varphi_S$ and $\varphi_T$, such as CEP instability, would average $I_{LO+NP}(\tau, \omega_d)$ to zero. On the other hand, in the balanced case, $I_{LO+NP}(\tau, \omega_d) \approx \varphi_T - \varphi_S$. When CEPs of the sampling and test pulses are locked, i.e. when pulses originated from the same laser, the CEP fluctuations are locked as well. In this way, phase fluctuations in the sampling and test pulses cancel each other out, enabling extremely sensitive detection without the need for CEP-stable pulses.

When $\varphi_T = \varphi_S = 0$, fieldoscopic measurement with CEP-stabilized and unstabilized pulses is expected to yield identical results. To confirm this, we set up a GHOST detection based on the balanced SHG/SFG channel ($m = n = 2$) utilizing a state-of-the-art octave-spanning (450 – 1000 nm), CEP-stable, sub-3 fs pulse [36], which is split into two orthogonally polarized replicas – sampling and test – with a controlled time delay τ between them. After setting the field strength and the CEP of each pulse, the pulses are recombined and collinearly focused on the nonlinear medium. The test pulse was kept weak while the field strength of the sampling pulse was approximately 1 V/Å. We use a z-cut α-quartz crystal of ~ 12 μm thickness as the nonlinear medium. The spectral response of the SHG/SFG channel covers the entire spectrum of the test pulse [30], while the CEP stability of the pulses can be switched off on demand. We first turn on the CEP stability and set the CEP of the sampling and test pulses to zero. Fig. 1b,c confirms that our balanced detection yields identical field traces whether or not the pulses are CEP-stabilized, as predicted.

To confirm that the measurement remains sensitive to the CEP of the test pulse, we vary its CEP by inserting a small amount of material using fused silica wedges (see Supplementary Information) and repeat the measurement. As can be seen in Fig. 1d, the change in the CEP of the test pulse is clearly captured by the measurement. We also note the enormous sensitivity of the measurement, which we attribute to the heterodyne detection.

**Fieldoscopy of strong (classical) coherent states**

After confirming that the $m = n$ channel provides a unique opportunity to access the electric field of light with CEP-unstabilized pulses, we test the concept directly with a laser oscillator. In this experiment we utilize commercial tabletop mode-locked Yb-based laser (Pharos, light conversion). In contrast to the previous measurements, which require sophisticated infrastructure for the generation of short (sub-3 fs) and intense (~ 0.4 mJ) few-cycle CEP-stable laser pulses at 1 kHz repetition rate, we employ a simple mode-locked laser delivering ~ 150 fs, CEP-unstable, ~ 8 nJ laser pulses centered at 1030 nm wavelength and with a 75 MHz repetition rate (see Supplementary Information for details). Fig. 2 shows the results of the experiment demonstrating that the fieldoscopic measurement previously feasible only with sophisticated infrastructure is now possible with simple table-top lasers.

We study the concept in detail by measuring the electric field of the laser oscillator at various test and sampling pulse energies (Fig. 2a). For each measurement, we measure a mean value of the signal and its standard deviation with a lock-in amplifier. Since our signal is expected to be proportional to the electric field of the test pulse, we expect a linear power scaling of the mean signal with the field strength of the test pulse – that is confirmed by the experiment (Fig. 2d). On the other hand, three photons of the sampling pulse contribute to the heterodyne detection in the SHG/SFG ($m = n = 2$) channel (Eq. 5). We conduct a similar power scaling experiment by scaling the field strength of the sampling pulse while keeping the test field unchanged (Fig. 2f).

In ideal environment with only quantum noise (due to photon statistics) and in the absence of classical noise (jitter, power source or environmental instability) one expects the standard deviation of the electric field amplitude to remain constant, regardless of the number of photons in the pulse [37]. In our experiment however, we observe that the standard deviation of the measurement scales linearly with the test field strength (Fig. 2e). This is expected since any practical system is a subject of the classical noise. We attribute this scaling to the dominance of classical noise over quantum noise in the regime of strong (classical) coherent states consisting of many photons. This suggests that in this regime, the standard deviation is primarily limited by experimental stability, rather than the probabilistic nature of the photon distribution. However, the contribution of the classical noise is expected to decrease as the test field (i.e., the number of photons) is decreased. Conversely, quantum noise – which is inherently associated with photon statistics – becomes more pronounced as the number of photons in the pulse is decreased. Therefore, we expect the quantum noise to be dominant in the regime of weak coherent states.

**Fieldoscopy of weak (quantum) coherent states**

Our approach not only allows us to utilize stable table-top lasers at high 75 MHz repetition rate, but it also removes the noise associated with the CEP fluctuations and highly non-linear light-matter interaction. Additionally, in contrast to broadband few cycle pulses (Fig. 1), narrow bandwidth laser oscillator pulses allow us to utilize thick nonlinear crystal (for higher signal strength) since the phase matching restriction is drastically eased. Having these advantages, we perform electric field measurement of very weak laser pulses to enter the regime of weak coherent states where quantum properties dominate over classical. Since our laser photon energy corresponds to about 1.2 eV (1030 nm wavelength) one enters this regime at pulse energies on the order of 193 zepto joules. We attenuate the test pulse down to the zepto joule level and perform the measurement. We measure the test field scaling as in Fig. 2d but in the zepto joule regime. Surprisingly, despite the fact that pulse energy is of zepto joules only, the sensitivity of the detection remains similar to the strong (classical) pJ-scale regime. As discussed earlier we mainly attribute this to the absence of the CEP noise and low-order non-linear light-matter interaction. However, stability of the table-top laser, MHz repetition rate, thick nonlinear medium and eased phase matching condition are important as well.

We note that in our classical regime, the measured signal and a corresponding standard deviation scale linearly with the test field strength (Fig. 2d,e). Since we are primarily interested in the deviation from the classical regime, in Fig. 3b,c we divide the mean and the standard deviation by the test field strength ($\sqrt{\langle n \rangle}$), so that in the classical regime these ratios are constant (which we further normalize to 1). Interestingly, when the mean photon number of the test pulse approaches 1 (~ 193 zJ), we observe a clear breakdown of the classical regime, both in the measured signal and the standard deviation. In particular, in the regime of weak coherent states, the scaling of the signal with the electric field becomes nonlinear. We note that in the classical description of electromagnetic waves, the scaling remains strictly linear regardless of the light field strength. The deviation from the linear scaling is a clear signature of the quantum nature of light, indicating that quantum properties (associated with coherence and photon statistics) become dominant and captured by our experiment.

**Photon statistics analysis**

Different quantum states of light are associated with unique photon statistics. In particular, coherent states follow the Poisson probability distribution, while non-coherent (thermal) light can be described by the Bose-Einstein probability distribution. We evaluate these distributions for the experimental mean photon numbers (Fig. 3d,e), with further details provided in Methods. As can be clearly seen, when the mean photon number $\langle n \rangle$ approaches 1, the probability of the test pulse containing zero photons exponentially increases in both, Poisson, and Bose-Einstein cases.

In contrast to intensity-based measurements, direct access to the electric field serves as an observable even in cases where the test pulse contains no photons (i.e., the vacuum state). This can be intuitively understood as follows. In intensity-based measurements, reducing the photon number by half (or blocking every other pulse) would halve the detected signal. In our case, reducing photon number by half reduces the field amplitude by only ~ $\sqrt{2}$ (since signal ∝ electric field, which scales with the square root of intensity), which is not equivalent to blocking every other pulse (see Supplementary Information for details).

Since our measurements are conducted with a laser, we expect the experiment to mainly follow the photon statistics of coherent light described by the Poisson probability distribution. We first model our experiment (see Methods) based on the Poisson probability distribution. Indeed, our model – where the electric field serves as an observable – predicts a deviation from the linear classical regime when the mean photon number approaches 1 and below (Fig. 3b,c). For the scaling of the mean value of the signal (Fig. 3b), our model agrees with the experiment. Surprisingly, however, the scaling of the experimental standard deviation significantly disagrees with the model. In particular, as the mean photon number approaches 1, the standard deviation is expected to increase (Fig. 3c, red curve) reaching its maximum when the mean photon number is 1. This is because for $\langle n \rangle = 1$ the probability of a pulses in the vacuum state is close to 50 % (Fig. 3d). Overall, the power scaling of the standard deviation based on Poisson statistics exhibits a distinct peak at $\langle n \rangle = 1$. In our experiment however, this feature is much less pronounced, which indicates a deviation from a perfectly coherent state.

To verify that the experimental observation is indeed a signature of the deviation from coherent properties, we re-model our experiment using the Bose-Einstein probability distribution instead of Poisson statistics (Fig. 3b,c, yellow curve). Interestingly, we find that the scaling of the mean value of the signal follows a similar trend to that predicted by Poisson statistics. This occurs because, although the Poisson and Bose-Einstein probability distributions describe different states of light, they appear similar in the regime of weak coherent states with $\langle n \rangle \approx 1$ (Fig. 3b,c, green and yellow bars). However, the scaling of the standard deviation according to the Bose-Einstein probability distribution differs significantly from that of Poisson statistics (Fig. 3c, yellow curve). In particular, the standard deviation monotonically decreases as the mean photon number decreases, which is similar to our experimental observation. Overall, our experiment appears to lie between of these two extremes – perfectly coherent and perfectly thermal states.

**Intrapulse coherence**

The observed deviation from perfectly coherent light could potentially be attributed to the variations in the coherence across the pulse. The direct access to the electric field in our experiment allows us to perform the same analysis for different regions of the pulse – the front, center, and tail. To achieve this, we utilize a Gabor method where a portion of the measured mean values and standard deviations of the pulse is multiplied by a Gaussian window as shown in Fig. 4a. The results of this analysis are presented in Fig. 4b,c. Regarding the mean signal value, we observe a trend similar to Fig. 3b, with no significant difference in scaling between different pulse regions. However, the standard deviation analysis reveals that different parts of the pulse exhibit different photon statistics. Notably, a characteristic feature of coherent Poisson statistics is clearly observed at the regions of the pulse with low intensity. As predicted by our model (Fig. 3c), the standard deviation peaks around $\langle n \rangle = 1$, a hallmark signature of a coherent state described by Poisson statistics. However, this feature is much less pronounced for intense parts of the pulse. Since most of the pulse energy is concentrated in this region, the overall pulse coherence is dominated by this region. This explains why the overall pulse coherence (Fig. 3b,c) looks similar to the coherence of the most intense part of the pulse. As previously discussed, variations in intrapulse coherence are generally expected in intense mode-locked lasers.

Finally, we model intrapulse coherence as a time-varying combination of coherent and thermal states of light, described by the Poisson and Bose-Einstein probability distributions (Fig. 4 d,e). We note a deviation from purely Poissonian statistics towards a mixed contribution of Bose-Einstein statistics, which increases with the number of photons in the pulse.

**Discussion**

The temporal dependence of the intrapulse coherence might have multiple origins. Although identification of the actual mechanisms requires further investigation, we attribute this deviation to an intensity-dependent decoherence mechanism (e.g., an elastic dephasing in the laser medium) in the laser — for example, due to fluctuations in the lasing transition or nonlinear coupling that effectively admixture a thermal component into the light field. In other words, the more intense parts of the pulse suffer slight decoherence. Such an effect could emerge from an elastic dephasing process $\propto \hat{a}^\dagger \hat{a}$ in the medium.

**Conclusion**

We introduced a novel approach for sampling of ultrafast light fields without the need for CEP-stabilized pulses. We benchmarked our method against a state-of-the-art CEP-stable detection, finding identical results, thereby validating its efficacy. The technique enables direct measurement of the light electric field with unprecedented sensitivity, reaching about 880 yoctojoules of average pulse energy. For our ~ 150 fs pulses, this corresponds to about 20 yoctojoules of average energy per optical cycle. We experimentally demonstrate the concept for pulse energies up to 1.24 pJ, achieving a dynamic range of more than 90 decibels. The high sensitivity allows measuring a time varying electric field of both, strong (classical) and weak (quantum) states of light. We have shown that the direct access to the electric field in the weak coherent state regime enables the study of intrapulse coherence and time-dependent photon statistics with sub-cycle temporal resolution.

By extending petahertz fieldoscopy into the quantum regime, our study provides a powerful new tool for investigating coherence at the most fundamental level. Freed from CEP-stabilization, our approach can be implemented with simple tabletop lasers, greatly broadening access to field-sensitive measurements. By bridging ultrafast science with quantum optics, our work provides a new platform for exploring quantum coherence and light-matter interactions at sub-cycle timescales. Future experiments could enable the direct measurement of nonclassical states of light, such as entangled photons and squeezed vacuum in a single photon regime, to gain deeper insight into quantum decoherence mechanisms. This could advance quantum-enhanced metrology, sensing, and optical quantum information processing, where control over coherence and photon statistics is essential. Beyond practical applications, our concept represents a new paradigm for exploring light–matter interactions at the quantum level, enabling tests of fundamental quantum optics with unprecedented sub-cycle precision.

# Main figure legends

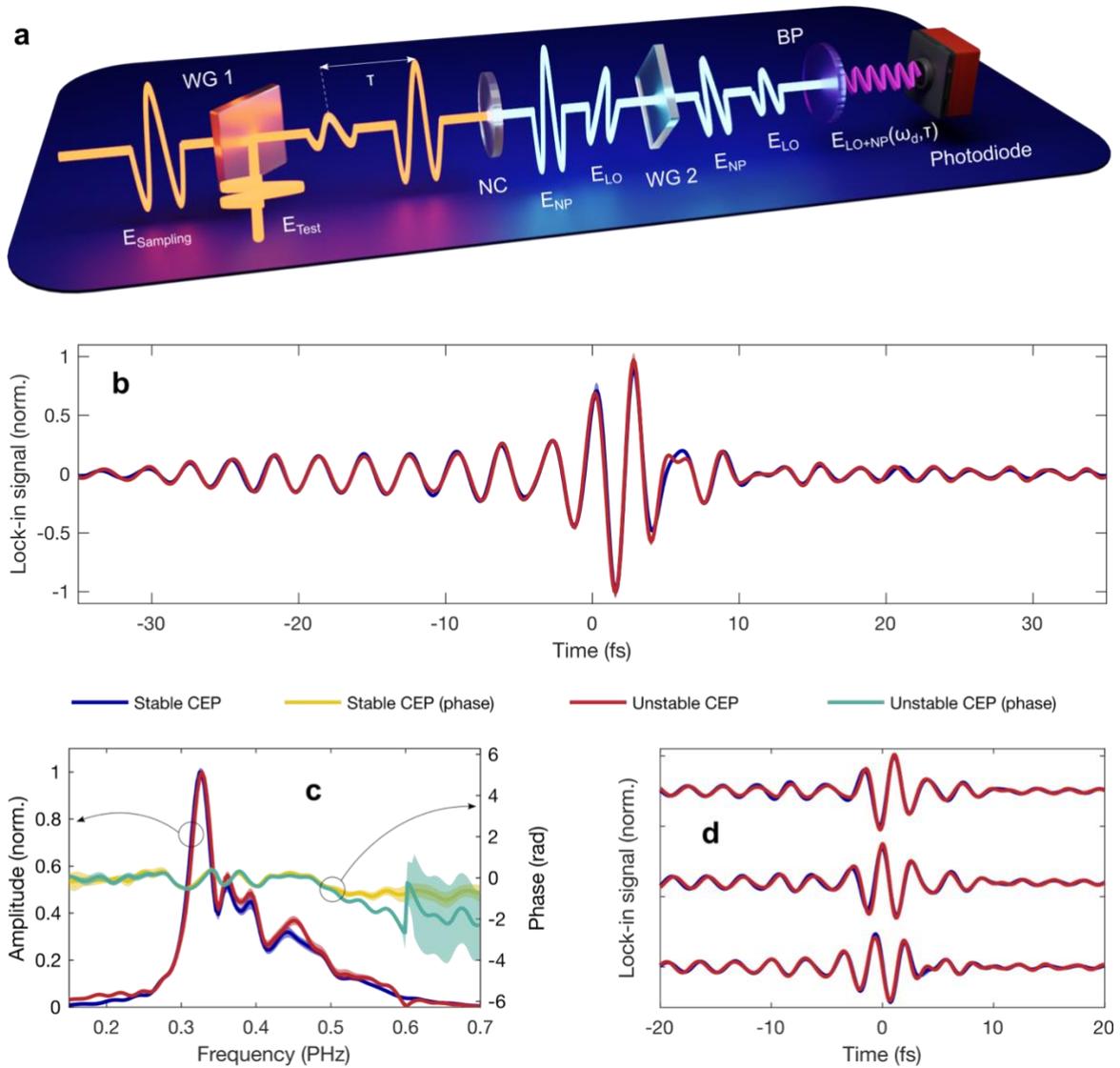

**Fig. 1 | Experimental concept. a**, Schematic of the optical setup. $E_{Sampling}$ and $E_{Test}$ denote the time-varying electric fields of the sampling and test pulses, respectively. $E_{LO}$ and $E_{NP}$ represent the electric fields of the nonlinearly generated local oscillator and the nonlinearly mixed pulses. NC, WG, BP and τ refer to the nonlinear crystal, wire-grid polarizer, bandpass filter and the time delay between sampling and test pulses respectively. **b**, Benchmarking of the measured GHOST signals with

CEP-stabilized and CEP-unstabilized pulses. **c**, Fourier-transform spectra of the signals in (**b**), illustrating the frequency content with and without CEP stabilization. **d**, Measured GHOST traces of the test pulse transmitted through glass with three slightly different thicknesses. The measurement highlights the signal's sensitivity to the carrier-envelope phase (CEP) of the test pulse. The medium's thickness was adjusted using a pair of fused silica wedges. The shaded areas in **b**, **c**, **d** correspond to one standard deviation of the measurement.

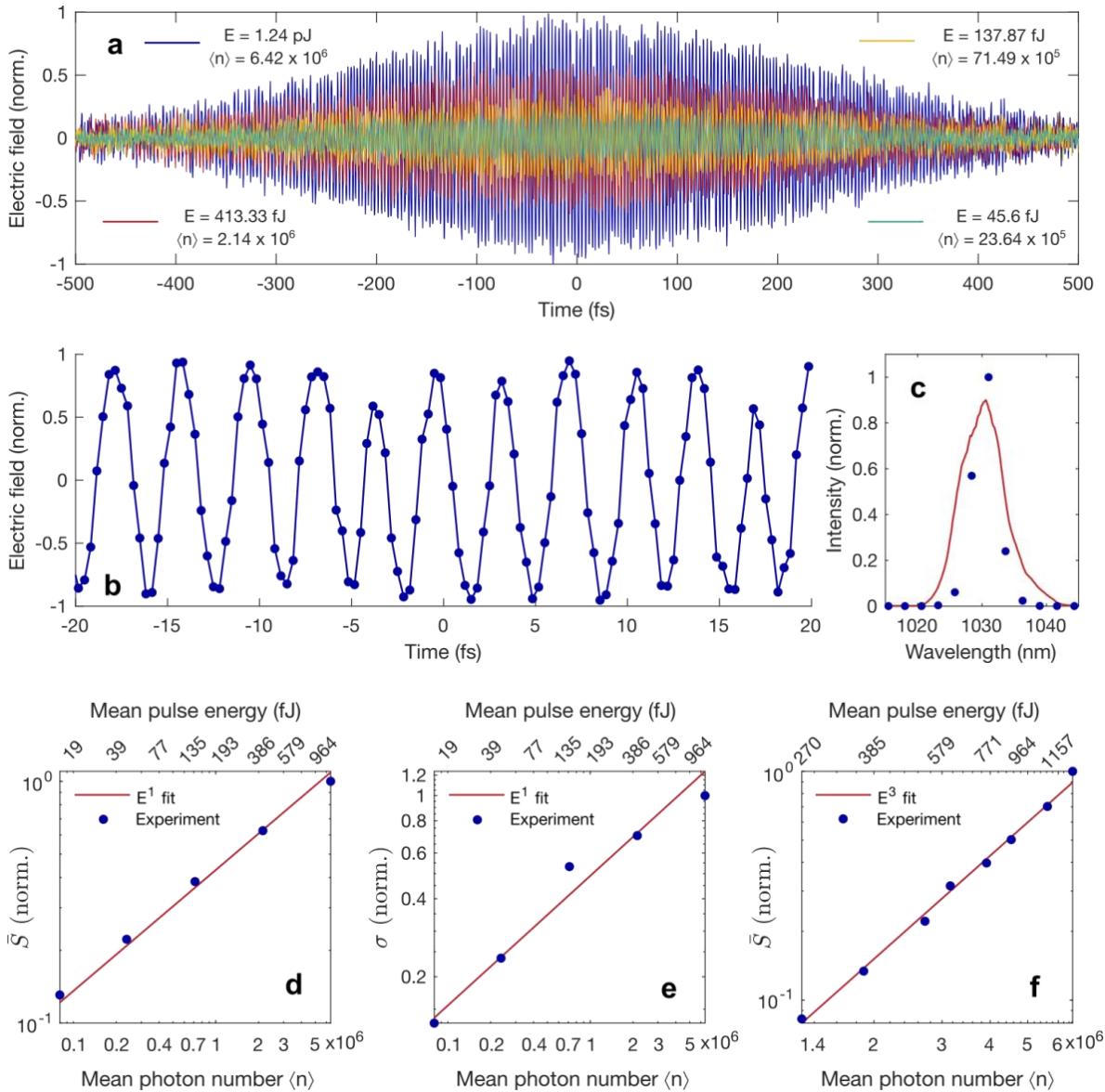

**Fig. 2 | Measurements of strong coherent states. a**, Measured electric fields of coherent states consisting of a large number of photons. **b**, Central part of the pulse ($\langle n \rangle$ = 6.42 x $10^6$) from the panel **a**, showing clear electric field oscillations with high fidelity. **c**, Benchmarking of the measured (blue dots) spectrum obtained by a Fourier transformation against a grating spectrometer (red curve). **d,** Dependence of the measured signal ($\bar{S}$) on the electric field strength of the test pulse. **e,** Dependence of the measured standard

deviation (σ) on the electric field strength of the test pulse. **f,** Dependence of the measured signal ($\bar{S}$) on the electric field strength of the sampling pulse.

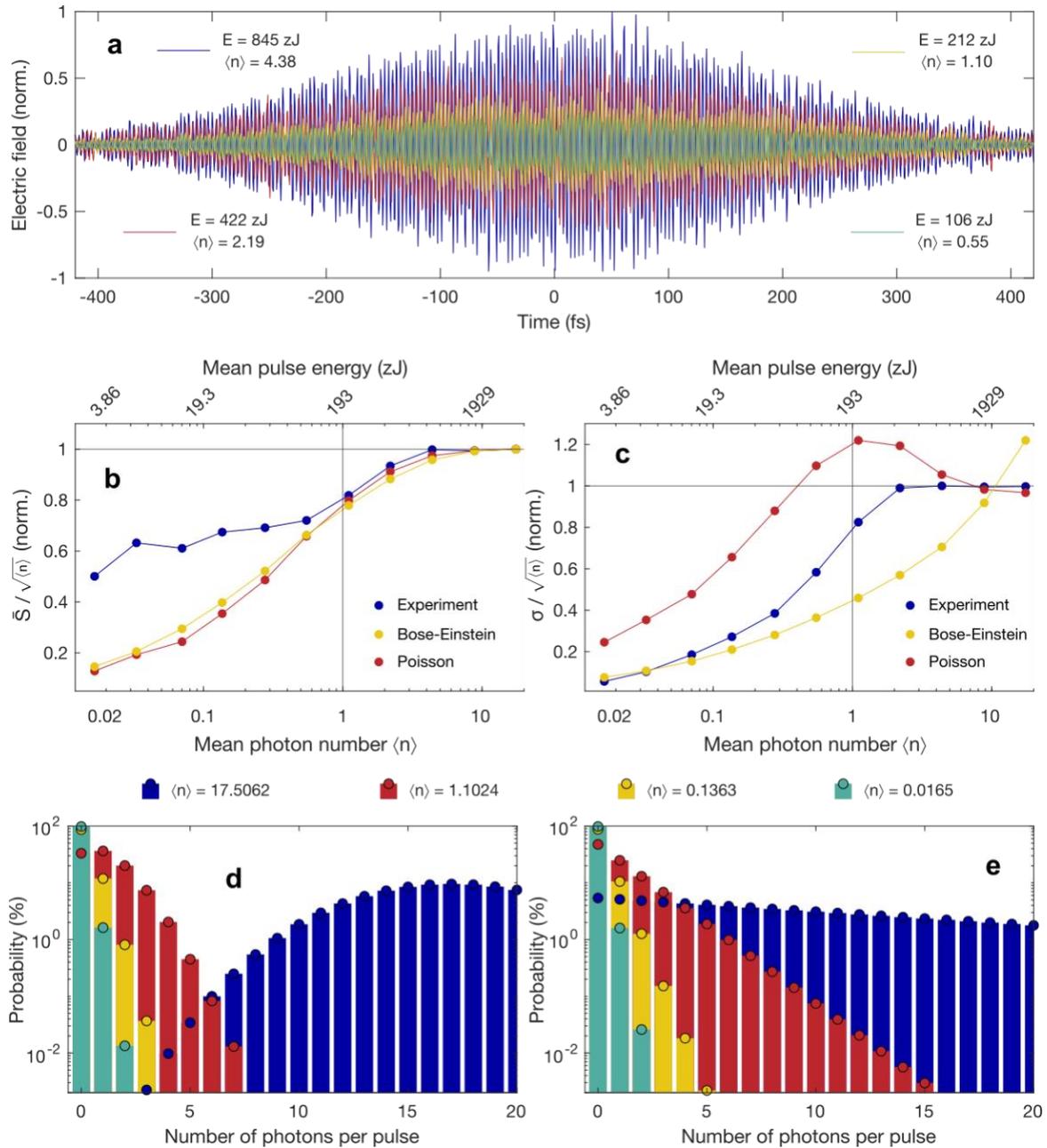

**Fig. 3 | Measurements of weak coherent states. a**, Measured electric fields of several weak coherent states consisting of a small number of photons. **b**, Measured and simulated breakdown of the mean signal ($\bar{S}$) from the linear scaling. **c**, Measured and simulated breakdown of the linear scaling of the measured standard deviation ($\sigma$). Poisson (**d**) and Bose-Einstein (**e**) probability distributions for mean photon numbers in (**a**).

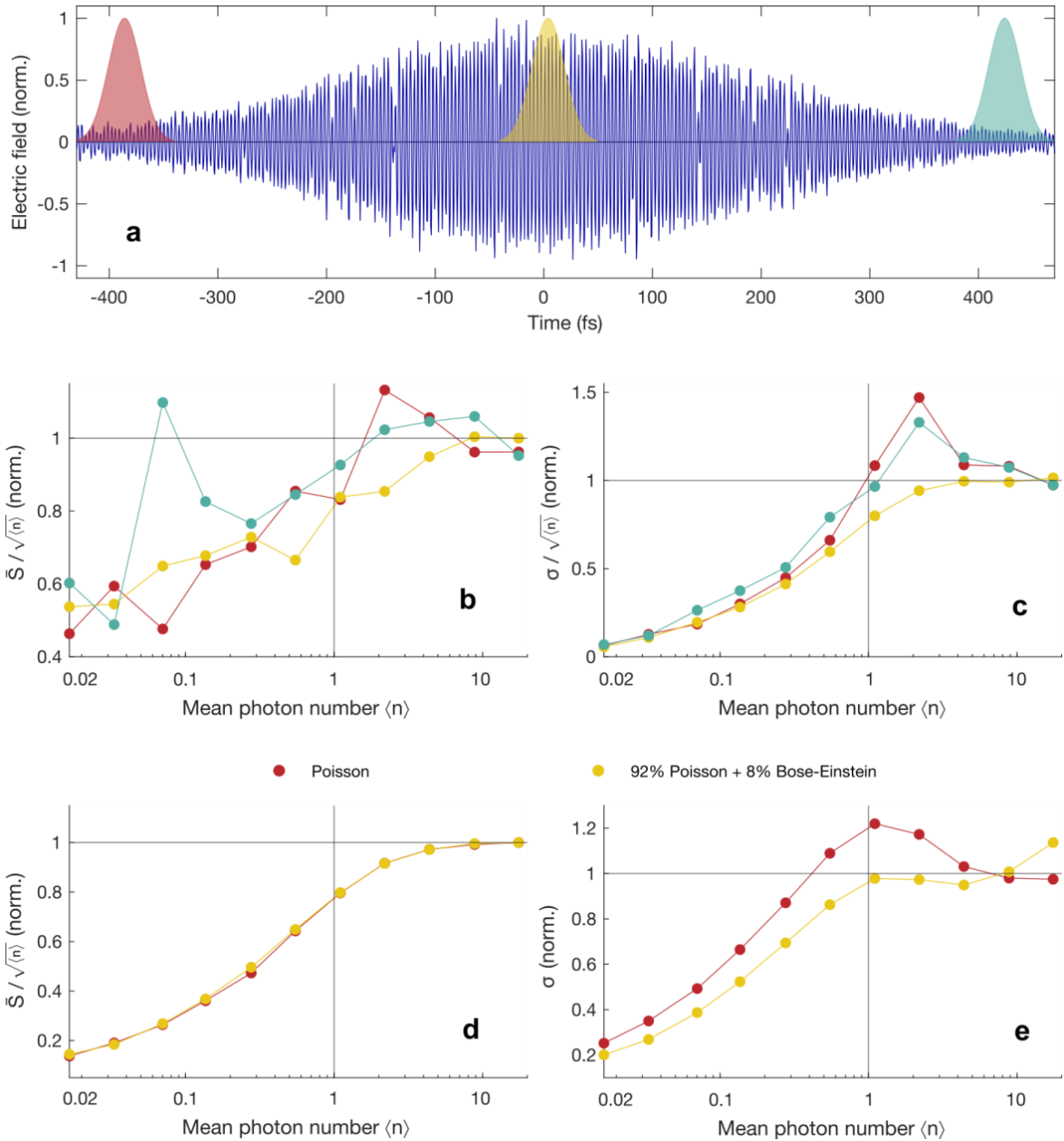

**Fig. 4 | Measurements of the intrapulse coherence. a**, Measured electric field of a weak coherent state ($\langle n \rangle$ = 4.38) with corresponding Gaussian windows used for the evaluation of intrapulse photon statistics in **b**, **c**. **b**, Measured breakdown of the linear scaling of the mean signal ($\bar{S}$). **c**, Measured breakdown of the linear scaling of the standard deviation (σ). **d**, Simulated breakdown of the linear scaling of the mean signal

($\bar{S}$). **e**, Simulated breakdown of the linear scaling of the standard deviation ($\sigma$). The colors of the curves in (**b**) and (**c**) correspond to the Gaussian windows in (**a**).

# Methods

**Theoretical model**

A coherent state of light is a quantum state of the electromagnetic field that can be described as the closest quantum analog to a classical light wave. It can be expressed in terms of Fock states (also known as photon number states), which are the eigenstates of the photon number operator

$$\hat{n} = \hat{a}^\dagger \hat{a}. \tag{1}$$

A coherent state $|\alpha\rangle$ is defined as an eigenstate of the annihilation operator $\hat{a}$:

$$\hat{a}|\alpha\rangle = \alpha|\alpha\rangle, \tag{2}$$

where $\alpha$ is a complex number representing the amplitude and phase of the state. The magnitude $|\alpha|^2$ is the mean photon number of the coherent state. The coherent state $|\alpha\rangle$ can be expanded as a superposition of Fock states $|n\rangle$ where $n$ is the number of photons in the state:

$$|\alpha\rangle = e^{-|\alpha|^2/2} \sum_{n=0}^{\infty} \frac{\alpha^n}{\sqrt{n!}} |n\rangle. \tag{3}$$

The probability of finding $n$ photons in the coherent state is governed by the Poisson distribution

$$P_{Poisson}(n) = \langle n|\alpha\rangle^2 = \frac{|\alpha|^{2n}}{n!} e^{-|\alpha|^2}, \tag{4}$$

with mean photon number $\langle n \rangle = |\alpha|^2$. The complex amplitude $\alpha = |\alpha|e^{i\varphi}$ encodes both the intensity ($|\alpha|^2$) and phase ($\varphi$) of the light. Similarly, the photon statistics of the non-coherent thermal light obeys Bose-Einstein probability distribution:

$$P_{Bose-Einstein}(n) = \frac{1}{(\langle n \rangle + 1)} \left(\frac{\langle n \rangle}{\langle n \rangle + 1}\right)^n. \tag{5}$$

In the Poisson distribution the variance in the photon number equals the mean $\Delta n^2 = \langle n \rangle = |\alpha|^2$, while the standard deviation $\sigma = \sqrt{n}$. In contrast, a thermal (Bose-Einstein) distribution has a variance larger than its mean (i.e., greater than $\langle n \rangle$), indicating larger number fluctuations than a coherent state.

In our experiment, the strong sampling pulse of 63 pJ pulse energy and with a photon energy of 1.2 eV ($1.9^{-19}$ Joules, 1030 nm wavelength) corresponds to $\langle n \rangle = 3.27*10^9$. The test pulse energies and corresponding mean photon numbers $\langle n \rangle$ utilized in the experiments are summarized in the table below.

| Energy | 3.19 | 6.37 | 13.54 | 26.28 | 53.35 | 105.91 | 212.61 | 422.83 | 845.65 | 1704 | 3376.2 | zJ |
|---|---|---|---|---|---|---|---|---|---|---|---|---|
| $\langle n \rangle$ | 0.0165 | 0.033 | 0.0702 | 0.1363 | 0.2766 | 0.5491 | 1.1024 | 2.1924 | 4.3848 | 8.8357 | 17.5062 | |

We statistically model the sampling pulse based on Eq. 4 with a mean photon number $\langle n \rangle = 1000000$, while the test pulse is modelled with $\langle n \rangle$ from the table above.

We assume that the photon statistics of the sampling pulse follows the Poisson distribution $P_{sampling}(n) = P_{Poisson}(n)$, while the test pulse is a linear combination of the Poisson and Bose-Einstein distributions: $P_{test}(n) = A * P_{Poisson}(n) + (1 - A) * P_{Bose-Einstein}(n)$.

The experiment involves three stages: 1) Second harmonic generation from the sampling pulse; 2) Sum-frequency generation between the sampling and test pulses; 3) Interference of the second harmonic and the sum-frequency fields.

For modeling of the second harmonic, two numbers of photons ($n_1$ and $n_2$) are randomly chosen from $P_{sampling}(n)$. Since in the second harmonic generation two fundamental photons combine to create a photon with higher energy, the number of second harmonic photons is limited either by $n_1$ or $n_2$. For instance, when $n_1 = 10$ and $n_2 = 20$, $n_{SHG} = 10$.

In the sum-frequency generation two numbers of photons ($n_1$ and $n_2$) are randomly chosen from $P_{sampling}(n)$ and $P_{test}(n)$. Similarly, to the second harmonic generation, in the sum-frequency generation, a photon from the sampling pulse and a photon from the test pulse combine to create a sum-frequency photon. The number of second harmonic photons is limited either by $n_1$ or $n_2$. For instance, when $n_1 = 10$ and $n_2 = 20$, $n_{SFG}$ is 10.

Finally, the measured signal due to the SHG/SFG interference is modeled as

$$S(n_{SHG}, n_{SFG}) \approx 2 * \sqrt{n_{SHG}} * \sqrt{n_{SFG}}, \qquad (6)$$

where $S$ is a measured signal. With this we conduct a Monte Carlo simulation where we repeat this procedure 10000 times to obtain a statistical distributing of $S$, from which we evaluate the mean ($\bar{S}$) and the standard deviation ($\sigma$).

**Benchmarking measurements with few-cycle CEP-stabilized pulses**

The laser beamline (more in SI) comprises a Ti:Sa oscillator (Rainbow 2, Spectra Physics), followed by chirped pulse amplification to 1 mJ pulse energy at 3 kHz repetition rate, further spectral broadening in a hollow-core fiber and a chirped mirror compressor. The experimental data acquisition (more in SI) was performed with a dual-phase lock-in amplifier (SR-830, Stanford Research Systems).

**Measurements with CEP-unstabilized oscillator pulses**

The laser beamline is a commercial tabletop mode-locked Yb-based laser (Pharos, light conversion).

# Acknowledgments

Authors acknowledge Ferenc Krausz and Max Planck Institute of Quantum optics for the experimental infrastructure and experiments with CEP-stabilized pulses.

# Funding

This study is supported by the Swiss National Science Foundation (SPF grant no. TMPFP2 217068, D.A.Z.). F.R. acknowledges funding from the Swiss National Science Foundation (Ambizione grant no. PZ00P2 186040).

# Author contributions

Conceptualization: D.A.Z. Methodology: D.A.Z. Experiments: D.A.Z., A.A. Theory: D.A.Z., F.R. Visualization: D.A.Z. Supervision: D.A.Z., A.R. Writing—original draft: D.A.Z. Review, scientific discussion, and editing: D.A.Z., A.A., F.R., A.R.

## Competing interests

Authors declare that they have no competing interests.

## Additional information

## Data availability

The datasets generated during and/or analyzed during the current study are available from the corresponding author upon reasonable request.

## Code availability

Programming scripts utilized for the data analysis and the theory are available from the corresponding author upon reasonable request.

## Corresponding author

Dmitry A. Zimin, email: dz366@cam.ac.uk

# Supplementary Materials for

# Fieldoscopy at the Quantum Limit

Dmitry A. Zimin[1,2], Arjun Ashoka[1], Florentin Reiter[3,4] and Akshay Rao[1]

[1]*Cavendish Laboratory, Department of Physics, Cambridge University, CB3 0HF Cambridge, United Kingdom*
[2]*Laboratory of Physical Chemistry, ETH Zürich, Vladimir-Prelog-Weg 2, 8049, Zürich, Switzerland*
[3]*Institute for Quantum Electronics, ETH Zürich, 8093 Zürich, Switzerland*
[4]*Quantum Center, ETH Zürich, 8093, Zürich, Switzerland*

## S1 Laser beamlines

The laser beamline for experiments was a commercial tabletop mode-locked Yb-based laser (Pharos, light conversion). The details on the experimental beamline with CEP-stabilized pulses can be found in [1, 2].

## S2 Optical schematics of experimental setups

The schematics of the experimental setups for CEP-stabilized and CEP-unstabilized measurements are shown on Figs. S1 and S2. For CEP-stabilized detection, the nonlinear medium was z-cut α-quartz crystal of ∼ 12 μm thickness [1]. For CEP-unstabilized detection, the nonlinear medium was 1 mm thick Barium Borate (BBO). The wedge pairs WP1 and WP2 were used to fine-tune the CEP in each optical arm.

## S3 Detection frequency

In both, CEP-stabilized and CEP-unstabilized experiments, the detection frequency was defined by the bandpass filter placed after the nonlinear crystal. In the CEP-stabilized case with a central wavelength of ∼ 750 nm, the narrow bandpass filter centered at 355 nm (Thorlabs) was used. In the CEP-unstabilized case with 1030 nm central wavelength, the narrow bandpass filter center at 515 nm (Thorlabs) was used.

**S4 Data acquisition**

In CEP-stabilized experiments, the signal was a current generated by a silicon photodiode. This current was first converted to voltage with further amplification by a transimpedance amplifier (DLPCA-200, FEMTO Messtechnik). The amplified voltage signal was then connected to a dual-phase lock-in amplifier (SR-830, Stanford Research Systems), triggered by an electrical signal synchronized with a half of the repetition rate of the laser. The measured signal from the lock-in amplifier was read by software on a computer via GPIB interface (National Instruments).

In CEP-unstabilized experiments with a laser oscillator, the signal was a current generated by a silicon photodiode. This current was first converted to voltage with further amplification by a transimpedance amplifier (DLPCA-200, FEMTO Messtechnik). The amplified voltage signal was then connected to a dual-phase lock-in amplifier (Zurich instruments), triggered by an electrical signal synchronized with a chopper installed in the test arm. The measured signal from the lock-in amplifier was read by software on a computer via USB interface.

**S5 Fieldoscopy versus intensity-based photon detection**

In contrast to intensity-based methods, fieldoscopy measures a signal proportional to the electric field, which scales with the square root of the number of incoming photons. This distinction has important implications. When detecting light intensity or photon counts, reducing the photon number by a factor of 2—such as by blocking every second pulse—leads to a corresponding halving of the signal. However, when detecting the electric field, halving the photon number decreases the signal only by a factor of $\sqrt{2}$, which is not equivalent to blocking alternate pulses. Fig. S3 illustrates this contrast by comparing intensity detection and electric field detection in the regime of perfectly coherent states, governed by Poisson statistics. In the simulation, we analyze how the signal power and its standard deviation scale with the mean photon number ⟨n⟩. Overall, the results in Fig. S3 show that intensity measurements alone do not reveal the transition from classical to quantum light, in contrast to the electric field.

**S6 Fieldoscopy in the yoctojoule regime**

The fieldoscopy measurement of the test field with a mean pulse energy in the yoctojoule regime was performed identically as the measurements with higher pulse energies. The test pulse was attenuated with a set of three natural density filters and wire grid polarizers for fine tuning of the pulse energy. The transmission of the natural density filters was characterized by measuring the incident and transmitted powers of the laser beam consisting of many photons. After the transmission of the laser pulse through the set of natural density filters and wire grid polarizers, the measured average pulse energy was 876 yocto joules which corresponds to the 0.0045 mean photon number $\langle n \rangle$. The results of the fieldoscopic measurement are shown in Fig. S4. We note that in the regime of $\langle n \rangle$ = 0.0045, considering perfectly coherent state described by the Poisson distribution, 99.55 % of pulse are in the vacuum state (no photons). Out of the remaining 0.45 % of pulses that contain photons, only 0.22 % consist of more than one photon and 99.78 % of pulses are in a single photon state. This statistical distribution leads to a very irregular signal (Fig. S5a). However, the signal is more pronounced in the spectral domain. By taking a Fourier transformation of the measured trace shown in Fig. S4a, the signal around 0.29 PHz (1030 nm) frequency (Fig. S4c) can be clearly observed. Similarly to the measurements at higher pulse energies we also record a standard deviation for each temporal delay between sampling and test pulses (Fig. S4b). In the absence of the test field, the standard deviation remains constant. In the presence of the test electric field, we expect a standard deviation to follow an envelope profile of the field. We indeed observe that the measured standard deviation is increasing as the temporal delay approaches 0 fs, and then decreases. This trend follows the envelope profile of the test field similarly to the measurements with larger photon numbers. Lastly, since the standard deviation is proportional to the absolute value of the electric field, in the frequency domain of the recorded temporal standard deviation, we expect spectral components around second harmonic (0.58 PHz) of the fundamental test pulse frequency (0.29 PHz). Indeed, Fig. S4d shows a distinct peak around 0.58 PHz.

Overall, our measurement demonstrates that in the extreme regime of weak coherent states with $\langle n \rangle$ = 0.0045, the measurement becomes very noisy due to irregularity of the photon arrivals. However, even in this regime, the distinct signatures of the test field are clearly observed.

# Extended data figure legends

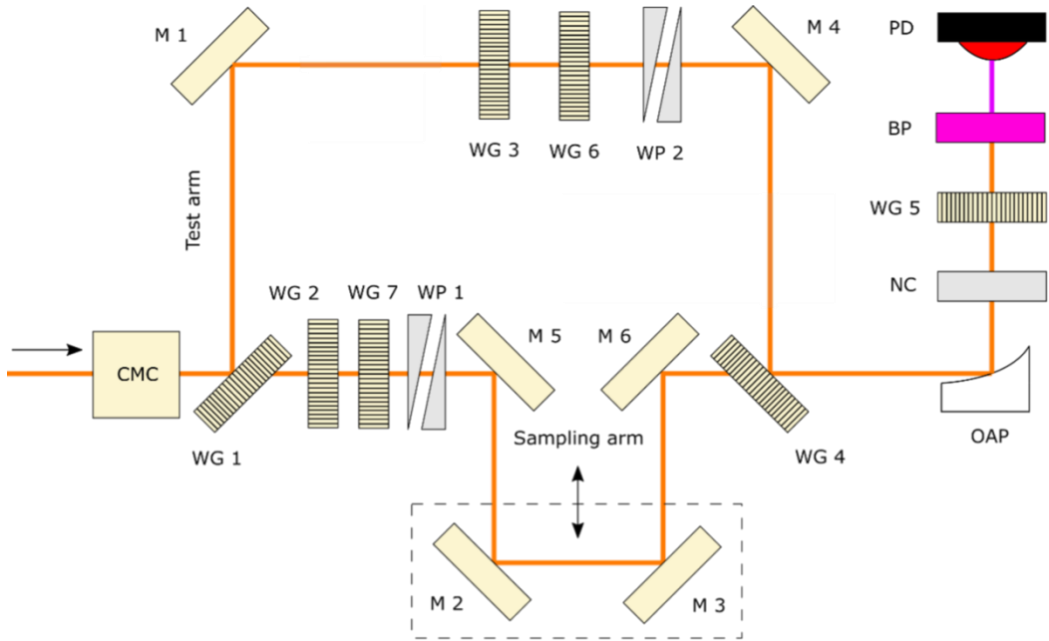

**Fig. S1 | Optical setup for CEP-stabilized experiments.** WG – wire-grid polarizer, WP – wedge pair, M – protected silver mirror, NC – nonlinear crystal, BP – bandpass filter, PD – photodiode, CMC – chirped mirror compressor, OAP – off-axis parabolic mirror.

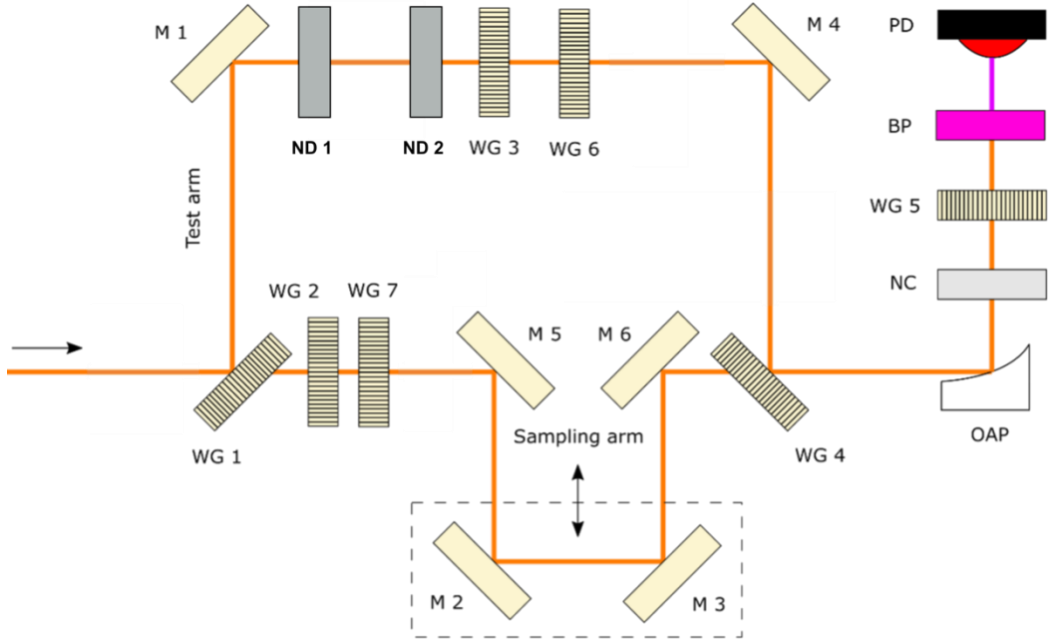

**Fig. S2 | Optical setup for CEP-unstabilized experiments with a laser oscillator.** WG – wire-grid polarizer, M – protected silver mirror, NC – nonlinear crystal, BP – bandpass filter, PD – photodiode, OAP – off-axis parabolic mirror, ND – natural density filter.

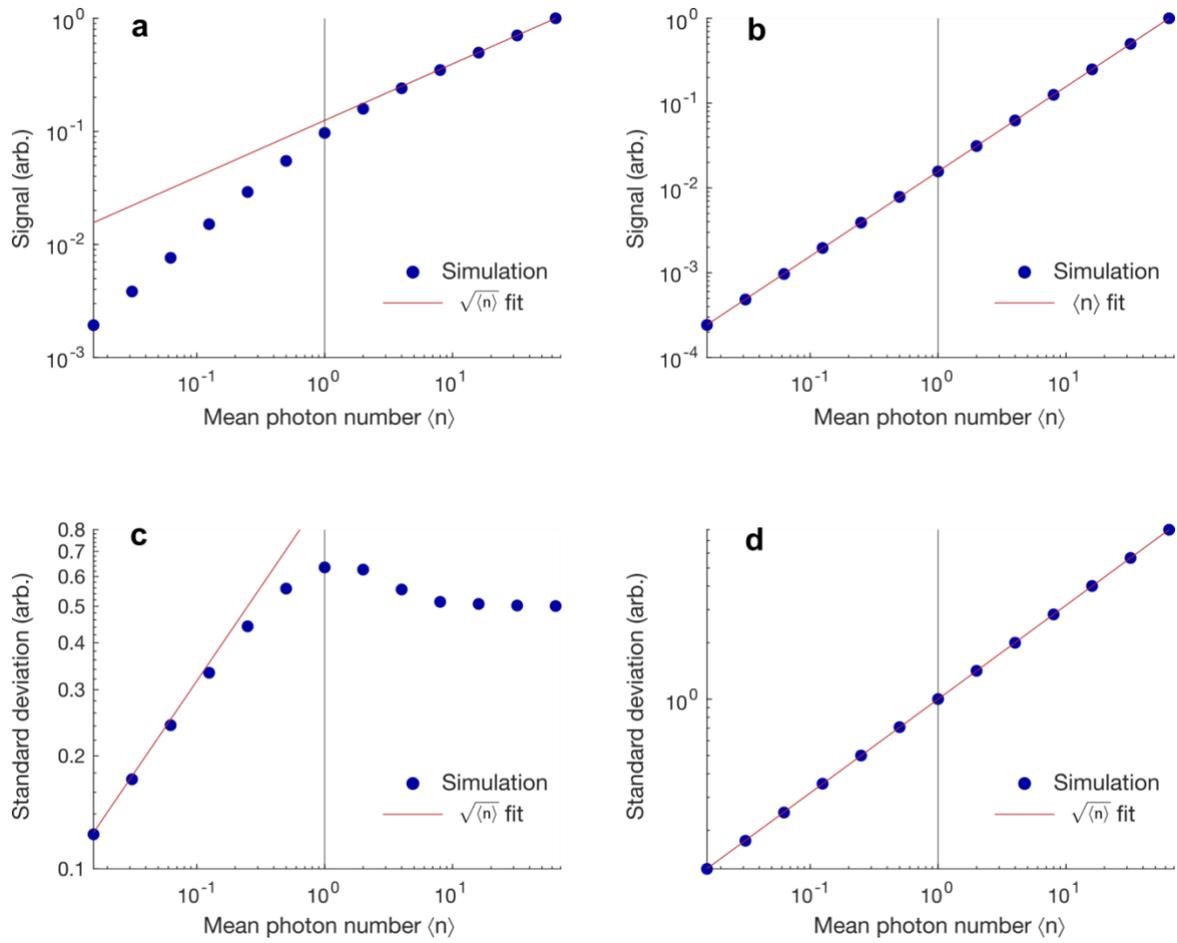

**Fig. S3 | Fieldoscopy vs spectroscopy. a**, Simulated field scaling vs mean photon number. **b**, Simulated intensity scaling vs mean photon number. **c**, Simulated scaling of the field standard deviation vs mean photon number. **d**, Simulated scaling of the intensity standard deviation vs mean photon number.

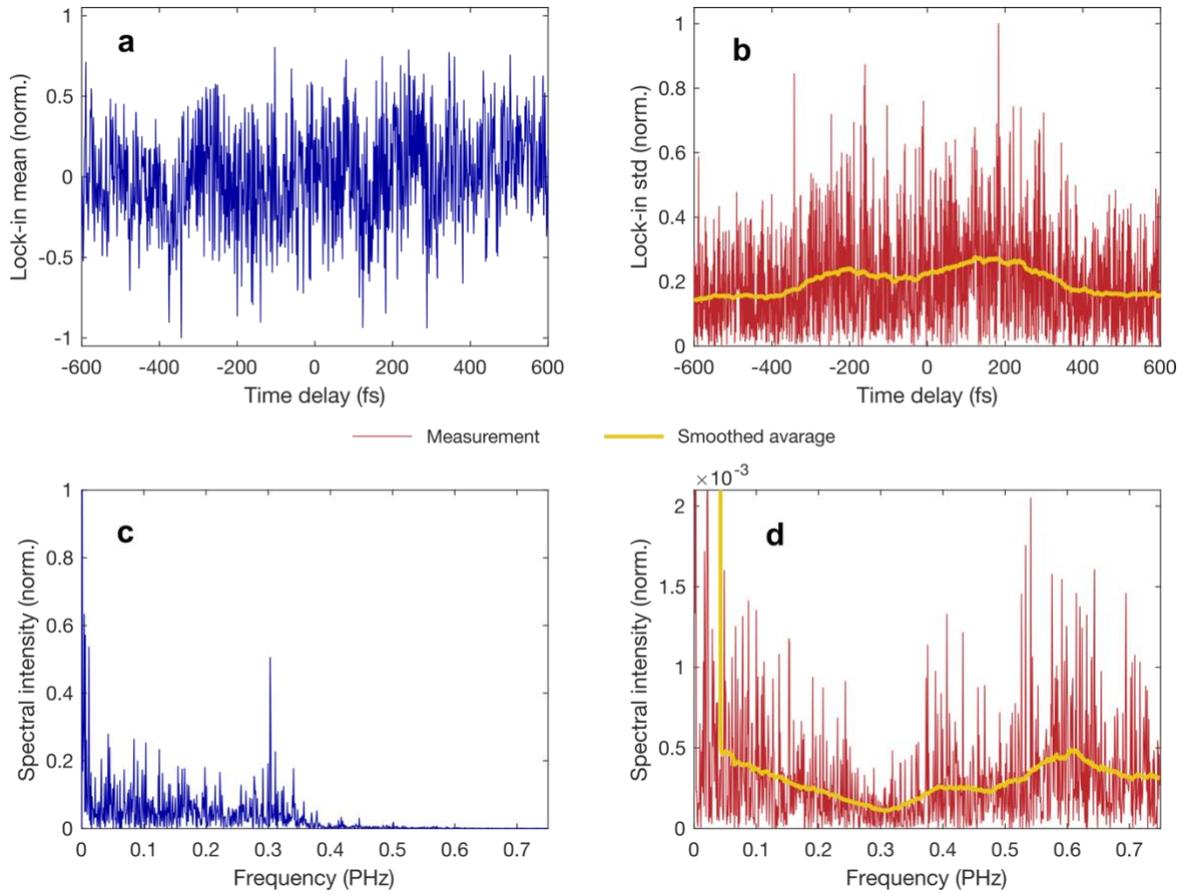

**Fig. S4 | Yoctojoule-level fieldoscopy. a**, Measured lock-in signal versus the time delay between sampling and test pulses. **b**, Measured standard deviation for each temporal delay in **a**. **c**, Spectrum of the measured signal in **a**. **d**, Spectrum of the measured standard deviation in **b**. The yellow curve represents a smoothed moving average of 200 points.

36